\documentclass[prc,aps,twocolumn, floatfix,nofootinbib,final,letterpaper]{revtex4}
\usepackage{epsfig}
\usepackage{amssymb}
\usepackage{amsmath}
\usepackage{amsfonts}
\usepackage{color, graphicx}
\usepackage{bm}% bold math

\def\lb{\langle}
\def\rb{\rangle}

\def\be{\begin{equation}}
\def\ee{\end{equation}}

\def\ad{a^\dagger_k}
\def\adb{a^\dagger_{\bar k}}
\def\a{a_k}

\def\alphd{\alpha^{\dagger}_k}
\def\alphdb{\alpha^\dagger_{\bar k}}
\def\alph{\alpha_k}
\def\alphb{\alpha_{\bar k}}
\def\phin{\varphi_n}
\def\Tr{{\rm Tr}}
\def\tr{{\rm tr}}

\newcommand{\pd}[2]{\frac{\partial #1}{\partial #2}}

\newcommand{\ket}[1]{\left | #1 \right >}
\newcommand{\bra}[1]{\left < #1 \right |}
\newcommand{\braket}[2]{\left< #1 | #2\right>}
\newcommand{\abs}[1]{\left | #1 \right |}
\newcommand{\avg}[1]{\lb #1 \rb}

\begin{document}
\title{Particle-number projection in the finite-temperature mean-field approximation
}
\author{P. Fanto$^1$, Y.~Alhassid$^{1}$, and G.F.~Bertsch$^{2}$ }
\affiliation{$^{1}$Center for Theoretical Physics, Sloane Physics
Laboratory, Yale University, New Haven, CT 06520\\
$^{2}$Department of Physics and Institute for Nuclear Theory,
Box 351560\\ University of Washington, Seattle, WA 98915}
\date{\today}

\begin{abstract}
Calculation of statistical properties of nuclei in a finite-temperature mean-field theory requires projection onto good particle number, since the theory is formulated in the grand canonical ensemble.  This projection is usually carried out in a saddle-point approximation. Here we derive formulas for an exact particle-number projection of the finite-temperature mean-field solution.  We consider both deformed nuclei, in which the pairing condensate is weak and the Hartree-Fock (HF) approximation is the appropriate mean-field theory, and nuclei with strong pairing condensates, in which the appropriate theory is the Hartree-Fock-Bogoliubov (HFB) approximation, a method that explicitly violates particle-number conservation. For the HFB approximation, we present a general projection formula for a condensate that is time-reversal invariant and a simpler formula for the Bardeen-Cooper-Schrieffer (BCS) limit, which is realized in nuclei with spherical condensates. We apply the method to three heavy nuclei: a typical  deformed nucleus $^{162}$Dy, a typical spherical nucleus $^{148}$Sm, and a transitional nucleus $^{150}$Sm in which the pairing condensate is deformed.  We compare the results of this projection with results from the saddle-point approximation and exact shell model Monte Carlo calculations.  We find that the approximate canonical HF entropy in the particle-number projection decreases monotonically to zero in the limit when the temperature goes to zero. However, in a nucleus with a strong pairing condensate, the approximate canonical HFB entropy in the particle-number projection decreases monotonically to a negative value, reflecting the violation of particle-number conservation.  Computationally, the exact particle-number projection is more efficient than calculating the derivatives required in the saddle-point approximation. 
\end{abstract}


\maketitle

\section{Introduction}

Finite-temperature mean-field approximations, in particular the finite-temperature Hartree-Fock (HF) and Hartree-Fock-Bogoliubov (HFB) approximations~\cite{Goodman1981,Tanabe1981}, are commonly used in the calculation of statistical properties of nuclei such as level densities~\cite{Hilaire2006}. These approximations are computationally efficient and therefore suitable for global studies of nuclear properties.

The nucleus is a finite-size system.  Consequently, the appropriate ensemble to describe it is the canonical ensemble with fixed numbers of protons and neutrons.  However, the finite-temperature HF and HFB approximations are formulated in the grand canonical ensemble through the variation of the grand thermodynamic potential. It is then necessary to carry out a reduction to the canonical ensemble to restore the correct proton and neutron numbers. This reduction is usually done  by expressing the canonical partition function as an inverse Laplace transform of the grand canonical partition function and evaluating this transform in the saddle-point approximation~\cite{BM1975,Kerman1981}.  A recent benchmarking of the accuracy of finite-temperature mean-field approximations identified significant problems with this saddle-point approach~\cite{Alhassid2016}. In particular, the method breaks down when the particle-number fluctuations are small.  This occurs in the HF approximation at low temperatures, and a better approximation in such a case is the discrete Gaussian approximation introduced in Ref.~\cite{Alhassid2016}. However,  in the low-temperature limit of the HFB, both the saddle-point and the discrete Gaussian approximations yield unphysical results. Specifically, the approximate canonical entropy becomes negative.   

Here we address the restoration of particle-number conservation using an exact particle-number projection~\cite{Ring1980,Rossignoli1994}. Particle-number projection can be combined with a mean-field approximation in two ways: (i) variation after projection (VAP), in which the variation to determine the mean-field solution is carried out in the projected subspace, and (ii) projection after variation (PAV), in which the projection operator is applied to the mean-field solution in the grand canonical ensemble.   In principle, VAP should result in a more accurate approximation.  It has been formulated for the zero-temperature HFB~\cite{Sheikh2000} and applied to even-even nuclei~\cite{Anguiano2001,Stoitsov2007}.  VAP has also been formulated at finite temperature in the Bardeen-Cooper-Schrieffer (BCS) approximation to calculate the pairing gaps in small model spaces~\cite{Gambacurta2012,Gambacurta2013}.  However, finite-temperature VAP requires the calculation of the entropy of the particle-number projected mean-field density operator. This is difficult to carry out in the presence of a pairing condensate since the density operator does not commute with the particle-number operator.  To the best of our knowledge, this method has not yet been applied to heavy nuclei at finite temperature.

In contrast, particle-number projection after variation is feasible at finite temperature in both the HF and HFB approximations.  A general formalism was presented in Ref.~\cite{Rossignoli1994}.  An alternate but equivalent formalism was derived in Ref.~\cite{Tanabe2005}.  Particle-number PAV has also been applied at finite temperature in the context of the BCS approximation~\cite{Esashika2005}.  However, in the HFB case, the general formalism includes a sign ambiguity.  Moreover, there has been no systematic assessment of the accuracy of particle-number PAV for heavy nuclei in finite-temperature mean-field theories.  

In this article, we derive unambiguous expressions for particle-number PAV for the HF and the HFB approximations and assess their accuracy numerically.  For brevity, we will refer to particle-number PAV as the particle-number projection (PNP). In the HF approximation, particle number is conserved and the PNP is straightforward~\cite{Ormand1994}.  However, the HFB approximation violates particle-number conservation at temperatures below the pairing transition for either protons or neutrons.  Assuming the HFB condensate is invariant under time-reversal symmetry, which is a property of most even-even nuclei, we derive a general formula for the PNP that is free of any sign ambiguity.  From this general expression, we derive a simpler expression for the PNP in the BCS limit of the HFB approximation.  We apply the PNP to even-even nuclei for three typical cases: (i) a strongly deformed nucleus with weak pairing ($^{162}$Dy), (ii) a spherical nucleus with strong pairing ($^{148}$Sm), and (iii) a transitional deformed nucleus with non-negligible pairing ($^{150}$Sm).  Following Ref.~\cite{Alhassid2016}, we benchmark the PNP results against those of the shell model Monte Carlo (SMMC) method~\cite{Lang1993,Alhassid1994,Alhassid2016a}.  The SMMC results are exact up to a controllable statistical error associated with the Monte Carlo sampling.  We also compare the PNP results with those of the discrete Gaussian approximations used in Ref.~\cite{Alhassid2016}, which were shown to improve on the usual saddle-point approximation.  We observe both a qualitative and quantitative improvement at low temperatures.  In the HF case, the PNP entropy does not show the unphysical oscillations that develop in the discrete Gaussian approximations. In the HFB case, the entropy in the PNP becomes negative at temperatures below the pairing transition, but its errors are less than those in the discrete Gaussian approximations.  Ultimately, the PNP is limited by deficiencies associated with the broken symmetries of the mean-field approximations.  In a deformed nucleus, the PNP results do not include the contributions of rotational bands, while in a nucleus with a strong pairing condensate, the PNP entropy becomes negative at low temperatures due to the intrinsic violation of particle-number conservation in the grand canonical theory.  We explain these limitations and suggest possible methods for addressing them.

The outline of this article is as follows.  In Sec.~\ref{PNP-HF}, we discuss a formula for the particle-number projected partition function of the HF density operator.  In Sec.~\ref{PNP-HFB}, we derive a general formula for the particle-number projected partition function of the HFB density operator assuming the condensate is time-reversal invariant. This formula simplifies in the BCS limit of the HFB approximation and, in particular, for a spherical condensate.  In Sec.~\ref{canonical-statistical}, we discuss how to calculate the approximate canonical thermal energy, entropy and state density from the projected partition functions. In Sec.~\ref{results}, we compare the PNP results with those from the discrete Gaussian approximations and the SMMC for the three heavy nuclei discussed above.  Finally, in Sec.~\ref{conclusion}, we summarize our findings and provide an outlook for future work.

\section{Particle-number projection in the Hartree-Fock approximation}\label{PNP-HF}

We assume a generic nuclear Hamiltonian in Fock space spanned by a set of $N_s$ single-particle orbitals with a one-body part described by the matrix $t$ and an anti-symmetrized two-body interaction $\bar v$
\be
\hat{H} = \sum_{ij}t_{ij}a^\dagger_{i}a_{j} + \frac{1}{4}\sum_{ijkl}\bar{v}_{ijkl}a^\dagger_{i}a^\dagger_j a_l a_k \;.
\ee

  The HF approximation is appropriate for nuclei with weak pairing condensates.  The single-particle HF Hamiltonian is given by $\hat H_{HF}=\sum_{ij} h_{ij}a^\dagger_i a_j$  where $h=t +\bar v \varrho$ and $\varrho$ is the one-body density matrix determined self consistently \cite{Ring1980}.  In the HF basis $\hat H_{HF} =  \sum_k  \epsilon_k c^\dagger_k c_k$, where $\epsilon_k$ are the single-particle HF energies of the single-particle HF orbitals $k$.  The grand canonical HF partition function is 
\be \label{HF-gc}
Z^{HF} = \Tr e^{-\beta (\hat{H}_{HF} -\mu \hat N-\lb \hat V\rb) } = e^{\beta\lb\hat{V}\rb}\prod_{k}\left[1 + e^{-\beta(\epsilon_{k} - \mu)}\right] \;,
\ee
where the trace is over the Hilbert space of all particle numbers and $\mu$ is the chemical potential determined by $\tr \varrho = N$. The subtraction of $\lb\hat V\rb = \tr\left(\varrho\bar v\varrho\right)/2$ corrects for the double counting of the two-body interaction, ensuring that $\avg{\hat{H}}_{HF} = -\partial \ln Z^{HF} /\partial \beta$.  

The particle-number projection operator $\hat{P}_{N}$ projects any many-particle state in Fock space onto the subspace of states with $N$ particles.  In a model space spanned by a finite number $N_s$  of single-particle orbitals, $\hat{P}_N$ is given by a discrete Fourier sum
\be 
\hat{P}_{N} = \frac{1}{N_s}\sum_{n = 1}^{N_{s}}e^{i\phin (\hat{N} - N)}\;,
\ee
where $\phin = 2\pi n/N_{s}$ are discrete quadrature points~\cite{Ormand1994}.   We define the particle-number projected HF partition function for $N$ particles by 
\be\label{HF-partition}
Z^{HF}_{N} = \Tr\left[\hat{P}_N e^{-\beta(\hat{H}_{HF} -\lb \hat V\rb)}\right] = \frac{e^{-\beta\mu N}}{N_{s}}\sum_{n = 1}^{N_s}e^{-i\phin N}\zeta_n^{HF}
\ee
where 
\be\label{z-HF}
\begin{split}
\zeta_n^{HF} & = e^{\beta\lb\hat{V}\rb}  \Tr\left[e^{i\phin \hat{N}}e^{-\beta(\hat{H}_{HF} - \mu \hat N)}\right] \\
& =  e^{\beta\lb\hat{V}\rb} \prod_{k}\left[1 + e^{-\beta(\epsilon_{k} - \mu)+i\phin}\right].
\end{split}
\ee
This formula follows from the fact that $[\hat{H}_{HF},\hat{N}] = 0$ for all temperatures.  
 In (\ref{HF-partition}), we include the factor $e^{-\beta\mu N}$  to cancel the contribution of  $e^{\beta \mu\hat{N}}$ in $\zeta_n$.  This equation holds for any value of $\mu$, but choosing a value of $\mu$ that is close to the chemical potential determined by $N$ particles ensures the numerical stability of the Fourier sum in (\ref{HF-partition}).  

\section{Particle-number projection in the Hartree-Fock-Bogoliubov approximation}\label{PNP-HFB}

For nuclei with strong pairing condensates, the appropriate mean-field theory is the HFB approximation.  The HFB Hamiltonian is~\cite{Ring1980}
\be\label{HFB-H}
\hat{H}_{HFB} = \sum_{ij}\left(h_{ij}a^\dagger_i a_j + \frac{1}{2} \Delta_{ij} a^\dagger_{i}a^\dagger_{j} - \frac{1}{2}\Delta^{*}_{ij} a_i a_j\right) \;,
\ee
where $h = t + \bar{v}\varrho$ is the density-dependent single-particle Hamiltonian, $\Delta_{ij} = \sum_{ijkl}\bar{v}_{ijkl}\kappa_{kl}/2$ is the pairing field, and $\kappa$ is the pairing tensor.  We can rewrite (\ref{HFB-H}) using a matrix notation 
\be\label{HFB-Hp}
\begin{split}
\hat H_{HFB} -\mu \hat N =  \frac{1}{2} (a^\dagger \, a )&\left( \begin{array}{cc} h-\mu  & \Delta \\ -\Delta^\ast &  -(h^\ast -\mu) \end{array}\right) \left( a \atop a^\dagger\right)  \\ 
&+\frac{1}{2} \tr \, (h-\mu) \;.
\end{split}
\ee
The $2 N_s \times 2 N_s$ matrix in (\ref{HFB-Hp}) can be diagonalized by a unitary transformation to the quasiparticle basis  $\alpha, \alpha^\dagger$~\cite{Ring1980}.  In this basis
\be \label{HFB-Hq}
\begin{split}
\hat{H}_{HFB} -\mu \hat N &=\frac{1}{2}  \sum_{k} (E_{k}\alpha^\dagger_{k}\alpha_{k} - E_{k} \alpha_k \alpha^\dagger_k)  +\frac{1}{2} \tr\,(h-\mu)  \\
& =  \sum_{k} E_{k}\alpha^\dagger_{k}\alpha_{k} + \frac{1}{2} \tr \, (h - \mu) -\frac{1}{2} \sum_k E_k  \;,
\end{split}
\ee
where $E_{k}$ are the HFB quasiparticle energies.  The grand canonical HFB partition function is then
\be
\begin{split}
Z^{HFB} & = \Tr e^{-\beta (\hat{H}_{HFB} -\mu \hat N -\lb \hat V \rb)}\\
& = e^{-\beta U_0} e^{\frac{\beta}{2} \sum_k E_k } \prod_{k}\left(1 + e^{-\beta E_{k}}\right) \;,
\end{split}
\ee
where the subtraction of  $\lb \hat V \rb = \tr \left(\varrho\bar v\varrho\right)/2 + \tr \left(\kappa^\dagger \bar v \kappa\right)/4$  accounts for the double counting of interaction terms, and
\be\label{U_0}
U_0 =   \frac{1}{2} \tr \, (h-\mu)   -\lb \hat V \rb \;.
\ee

As in the HF case, we define the particle-number projected HFB partition function for $N$ particles by
\be\label{HFB-partition}
\begin{split}
Z^{HFB}_N & = \Tr \left[\hat{P}_{N} e^{-\beta(\hat{H}_{HFB} - \lb \hat V \rb)}\right] \\
&= \frac{e ^{-\beta\mu N} }{N_{s}}\sum_{n =1}^{N_s} e^{-i\phin N}\zeta_n^{HFB}
\end{split}
\ee
where
\be\label{z-HFB}
\zeta^{HFB}_n =   e^{\beta\lb\hat{V}\rb} \Tr\left[ e^{i\phin \hat{N}}e^{-\beta (\hat{H}_{HFB} - \mu \hat N)}\right]\;.
\ee
In contrast to the HF approximation, $[ \hat H_{HFB}, \hat N] \neq 0$.  Consequently, it is more difficult to evaluate the trace in (\ref{z-HFB}) than it was to evaluate the trace in (\ref{z-HF}).  In Sec.~\ref{tr-condensate} we derive a formula to evaluate this trace for an HFB Hamiltonian that is invariant under time reversal.  In Sec.~\ref{sph-condensate} we show that this formula simplifies considerably in the BCS limit of the HFB, in which the Bogoliubov transformation mixes any particle state with only its time-reversed counterpart.  

\subsection{A pairing condensate with time-reversal symmetry}\label{tr-condensate}

Here we assume that the HFB Hamiltonian is invariant under time reversal, and thus its quasiparticle states come in time-reversed pairs $\ket{k},\ket{\bar{k}}$ with degenerate energies $E_k =E_{\bar k}$.\footnote{If the condensate is also axially symmetric, the magnetic quantum number $m$, which is the projection onto the symmetry axis, is a good quantum number.  In this case, we choose the $k$ states to have positive signature, i.e.~$m = 1/2, -3/2, 5/2, -7/2, ...$, and $\bar k$ states have negative signature, i.e.~$- m$.}
The Bogoliubov transformation that defines the quasiparticle basis can then be fully expressed by
\be \label{Bogol-tr}
\left(\begin{matrix}\alph \\ \alphdb\end{matrix}\right) = \mathcal{W}^\dagger \left(\begin{matrix} \a \\ \adb\end{matrix}\right)
\ee
where $k$ runs over half the number of single-particle states from $1,...,N_{s}/2$.  In the following, we denote these states by $k >0$.  $\mathcal{W}$ is an $N_s\times N_s$ unitary matrix of the form
\be\label{mat-W}
\mathcal{W} = \left(\begin{matrix} \mathcal{U} & - \mathcal{V} \\ \mathcal{V} & \mathcal{U} \end{matrix}\right) \;.
\ee
The general Bogoliubov transformation matrix is $2N_{s}\times2N_s$-dimensional \cite{Ring1980}.  Thus, time-reversal symmetry reduces its size by a factor of 2. Using $E_k =E_{\bar k}$, the HFB Hamiltonian in (\ref{HFB-Hq}) can be rewritten as
\be\label{HFB-Htr}
\begin{split}
\hat{H}_{HFB} -\mu \hat N  & = \sum_{k > 0} E_k  (\alphd \alph + \alphdb\alphb)+\frac{1}{2}\tr\, (h - \mu) - \sum_{k>0}  E_k \\
& = \sum_{k>0} E_k (\alphd \alph - \alphb\alphdb) + \frac{1}{2}\tr\, ( h - \mu) \;. 
\end{split}
\ee
In a more compact notation,
\be\label{HFB-mtx}
\hat{H}_{HFB} -\mu \hat N  =  \xi^\dagger\mathcal{E}\xi + \frac{1}{2}\tr\, (h - \mu) \;,
\ee
where $\xi^\dagger = \left(\alpha^\dagger_{k_1},...,\alpha^\dagger_{k_{N_{s}/2}}, \, \alpha_{\bar{k}_{1}}, ..., \alpha_{\bar{k}_{N_{s}/2}}\right)$ and
\be\label{mat-E}
\mathcal{E} = \left(\begin{matrix} E & 0 \\ 0 & - E\end{matrix}\right) \;.
\ee
 The matrix $E$ is the diagonal matrix of the HFB quasiparticle energies $E_k$ ($k =1,\ldots, N_{s}/2)$.  Similarly, the number operator $\hat N$  can be written as
\be\label{N-mtx}
\begin{split}
\hat{N} & = \sum_{k>0}  \left(a_k^\dagger a_k +a_{\bar k}^\dagger a_{\bar{k}} \right) = \sum_{k>0} \left(a^\dagger_ka_k - a_{\bar k}a^\dagger_{\bar k}\right) + \frac{N_{s}}{2}\\
 & = \xi^\dagger \left(\mathcal{W}^\dagger\mathcal{N}\mathcal{W} \right) \xi + \frac{N_{s}}{2}  \;,
 \end{split}
\ee
where  
\be\label{mat-N}
\mathcal{N} = \left(\begin{matrix} 1 & 0 \\ 0 & -1 \end{matrix}\right) \;,
\ee
and where we have used the transformation in (\ref{Bogol-tr}).

Using (\ref{HFB-mtx}) and (\ref{N-mtx}), we can rewrite (\ref{z-HFB}) as
\be\label{z1-HFB}
\zeta^{HFB}_n =
 e^{-\beta U_0 } e^{i \phin N_s/2}\Tr\left[e^{i\phin \xi^\dagger \left( \mathcal{W}^\dagger\mathcal{N}\mathcal{W}\right)\xi}e^{-\beta\xi^{\dagger}\mathcal{E}\xi}\right].
\ee
To evaluate the trace, we use the group property of the exponentials of one-body fermion operators written in quadratic form.  This property states that the product of two such group elements is another group element
\be
e^{\xi^\dagger {\cal A} \xi} e^{\xi^\dagger {\cal B} \xi}  = e^{\xi^\dagger {\cal C} \xi}  \;,
\ee
where the matrix ${\cal C}$ is determined from the single-particle representation of the group
\be
e^{\cal A} e^{\cal B} = e^{\cal C} \;.
\ee
Applying this property to (\ref{z1-HFB}), we can rewrite it in the form
\be
\zeta^{HFB}_n = (-)^n e^{-\beta U_0}  \Tr\left[  e^{\xi^\dagger\mathcal{C}_n(\beta)\xi} \right]\;,
\ee
where the matrix ${\cal C}_n(\beta)$ is determined from
\be \label{expcn}
e^{\mathcal{C}_n(\beta)} = e^{i\phin \mathcal{W}^\dagger\mathcal{N}\mathcal{W}}e^{-\beta\mathcal{E}} = \mathcal{W}^\dagger e^{i\phin\mathcal{N}}\mathcal{W}e^{-\beta\mathcal{E}} \;.
\ee
Using the formula for the trace of the exponential of a one-body fermionic operator (see Appendix I), we find
\be \label{genzn}
\zeta^{HFB}_n = (-)^n e^{-\beta U_0}\det\left(1 + e^{\mathcal{C}_n(\beta)}\right) \;.
\ee
Using (\ref{expcn}), we obtain the final expression
\be\label{z-HFB-tr}
\zeta^{HFB}_n = (-)^n e^{-\beta U_0}\det\left(1 +  \mathcal{W}^\dagger e^{i\phin\mathcal{N}}\mathcal{W}e^{-\beta\mathcal{E}} \right) \;,
\ee
where $U_0$ is given by Eq.~(\ref{U_0}) and the matrices $\mathcal{W}$, $\mathcal{E}$ and $\mathcal{N}$ are given, respectively, by Eqs.~(\ref{mat-W}), (\ref{mat-E}) and (\ref{mat-N}). 

Eq.~(\ref{z-HFB-tr}) is a general formula that applies to any condensate with time-reversal symmetry.  In the HF case, the Bogoliubov transformation reduces to a unitary transformation among the particle basis operators, and $\mathcal{V}$ vanishes.   A formula valid for the most general case can be derived in a similar fashion by using $\xi^{\prime\dagger} = \left(\begin{matrix}\alphd &  \alph & \alphdb & \alphb \end{matrix}\right)$ (where $k = 1,...,N_{s}/2$) and making the dimension of the relevant matrices $2N_{s} \times 2N_{s}$~\cite{Rossignoli1994}.  However, the final expression for $\zeta^{HFB}_n$, given in Eq.~(3.46) of Ref.~\cite{Rossignoli1994}, involves a square root of a determinant.  This square root leads to a sign ambiguity that is difficult to resolve.  The method discussed here eliminates this sign ambiguity completely for the case when the condensate is time-reversal invariant by working with matrices of reduced dimension $N_s\times N_s$.

Eq.~(\ref{z-HFB-tr}) becomes numerically unstable at large $\beta$.  The reason for this instability can be seen in Eq.~(\ref{expcn}).  At large $\beta$, the diverging scales in the diagonal matrix $e^{-\beta\mathcal{E}}$ will dominate the smaller scales in the matrix product $\mathcal{W}^\dagger e^{i\phin\mathcal{N}}\mathcal{W}$.  We stabilize the calculation by the method discussed in Appendix II.

\subsection{The BCS Limit}\label{sph-condensate}
Eq.~(\ref{z-HFB-tr}) simplifies in the BCS limit, in which the quasiparticle representation mixes particle state $k$ with only its time-reversed counterpart $\bar k$
\be
\alph = u_k \a - v_k\adb\;;\qquad \alphdb = u_k\adb + v_k\a\;,
\ee
where $u_k$ and $v_k$ are real numbers satisfying $u_k^2 + v_k^2 = 1$. In this case, the Bogoliubov transformation in (\ref{Bogol-tr}) is described by diagonal matrix $\mathcal{U}$  and an anti-diagonal $\mathcal{V}$, which can be decomposed into a set of $N_{s}/2$ transformations 
\be\label{BCS-tr}
\left(\begin{matrix}\alph \\ \alphdb \end{matrix}\right)= \left(\begin{matrix} u_k & -v_k \\ v_k & u_k\end{matrix}\right)\left(\begin{matrix}\a \\ \adb\end{matrix}\right)
\ee
for each $\{k,\bar k\}$ pair. Eq.~(\ref{z-HFB-tr}) can then be rewritten as a product of the block determinants
\be\label{z-HFB-bcs}
\zeta_n^{HFB} = (-)^n e^{-\beta U_0}\prod_{k > 0}\det\left(1 + e^{i\phi_n\mathcal{W}^\dagger \mathcal{N}\mathcal{W}}|_k e^{-\beta \mathcal{E}}|_k \right)\;,
\ee
where 
\be
e^{i\phi_n\mathcal{W}^\dagger \mathcal{N}\mathcal{W}}|_k = \left(\begin{matrix}u_k^2 e^{i\phi_n}+ v_k^2 e^{-i\phi_n} & u_k v_k (e^{i\phi_n} - e^{-i\phi_n}) \\ u_k v_k (e^{i\phi_n} - e^{-i\phi_n}) & v_k^2 e^{i\phi_n} + u_k^2 e^{-i\phi_n}\end{matrix}\right)
\ee
and 
\be
e^{-\beta \mathcal{E}}|_k = \left(\begin{matrix} e^{-\beta E_k} & 0 \\ 0 & e^{\beta E_k}\end{matrix}\right)\;.
\ee
 Evaluating the determinants in (\ref{z-HFB-bcs}) we find the final expression
\be\label{z1-HFB-bcs-final}
\begin{split}
\zeta_n^{HFB} = e^{-\beta U_0}\prod_{k > 0}e^{\beta E_k}[u_k^2 & + e^{2i\phi_n}v_k^2 + 2e^{-\beta E_k + i\phi_n}  \\
&  + e^{-2\beta E_k}(v_k^2 + e^{2i\phi_n} u_k^2)]\;,
\end{split}
\ee
where we have used $\prod_{k > 0} e^{-i\phin} = e^{i\phin N_{s}/2} = (-)^n$.

The result in (\ref{z1-HFB-bcs-final}) can also be derived by writing explicitly the matrix elements of $e^{i\varphi\hat{N}}$  in the subspace spanned by the four many-body states $\ket{}_k = (u_k + v_k \ad\adb)\ket{}$, $\alphd\ket{}_k$, $\alphdb\ket{}_k$, and $\alphd\alphdb\ket{}_k$, and by evaluating the traces of $e^{i\phi_n\hat{N}}e^{-\beta \hat{H}_{HFB} - \mu \hat{N} - \avg{\hat{V}}}$ in each of these subspaces.

We note that Eq.~(\ref{z1-HFB-bcs-final}) is identical to Eq.~(25) of Ref.~\cite{Esebbag1993}.  A key difference between our work and the work in Ref.~\cite{Esebbag1993} is that we obtain Eq.~(\ref{z1-HFB-bcs-final}) as the special limit of a more general formula.  

An important case of the BCS limit is that of a spherical condensate, in which both the total angular momentum quantum number $j$ and the magnetic quantum number $m$ are preserved by the Bogoliubov transformation.  Consequently, the quasiparticle energies for a given $j$ are independent of $m$ and therefore are each $2j+1$-fold degenerate.  If the single-particle model space does not include more than one shell with the same $j$, the Bogoliubov transformation for this case is of the form (\ref{BCS-tr}), where $\{k,\bar k\}$ represent the states $\ket{k} = \ket{jm}$ and $\ket{\bar k} \propto \ket{j-m}$, with $m = 1/2,-3/2, 5/2 ,-7/2,...$ being the positive-signature states.  The parameters $u_k = u_j$ and $v_k = v_j$ depend only on $j$.  Therefore, Eq.~(\ref{z1-HFB-bcs-final}) simplifies to
\be\label{z-HFB-sph}
\begin{split}
\zeta_n^{HFB} = e^{-\beta U_0}\prod_j e^{\beta(j+\frac{1}{2})E_j}[ & u_j^2 + e^{2i\phi_n}v_j^2 + 2e^{-\beta E_j + i\phi_n} \\
& + e^{-2\beta E_j}(v_j^2 + e^{2i\phi_n} u_j^2)]^{j+\frac{1}{2}}\;.
\end{split}
\ee

\section{Canonical energy, entropy and level density in the mean-field approximation}\label{canonical-statistical}

The state density $\rho(E)$ at energy $E$ for a nucleus with proton and neutron numbers $N_p$, $N_n$ is an inverse Laplace transform of the canonical partition function $Z_c(\beta)$.  The average state density is obtained in the saddle-point approximation 
\be \label{sden}
\rho(E)  = \frac{1}{2\pi i}\int_{-i\infty}^{i\infty} d\beta\, e^{\beta E}Z_{c}  \approx \left(2\pi\left|\pd{E}{\beta}\right|\right)^{-1/2}e^{S_{c}(\beta)} \;.
\ee
Here $\beta$ is determined as a function of the energy $E$ by the saddle-point condition
\be \label{spcond}
E = -\pd{\ln Z_{c}}{\beta}  = E_c \;,
\ee
and  $S_c$ is the canonical entropy
\be \label{canentropy}
S_{c} = \ln Z_c +  \beta E_c \;.
\ee 
When $Z_c(\beta)$ is the exact canonical partition function, i.e., $Z_c(\beta) = \Tr (\hat P_N e^{-\beta \hat H})$ where $\hat P_N = \hat P_{N_p} \hat P_{N_n}$, the r.h.s. of Eq.~(\ref{spcond}) is the canonical energy $E_c=\Tr (\hat P_N e^{-\beta \hat H} \hat H)/ Z_c(\beta)$ and the entropy (\ref{canentropy}) is the exact canonical entropy $S_c=-\Tr (\hat D_N \ln \hat D_N)$ of the correlated density matrix $\hat D_N= \hat P_N e^{-\beta\hat H} /Z_c(\beta)$.

In the particle-number projected finite-temperature HF approximation, the proton (neutron) partition function $Z_{N_{p(n)}}^{HF}$ is given by Eqs.~(\ref{HF-partition}) and (\ref{z-HF}) and the approximate canonical HF partition  is $Z^{HF}_c = Z_{N_p}^{HF}Z_{N_n}^{HF}$.  We define the approximate canonical HF energy $E_c^{HF}$  by the saddle-point condition
\be\label{can-E}
E= -\pd{\ln Z^{HF}_c}{\beta} =  E_c^{HF}\;.
\ee
Similarly, we define the approximate canonical HF entropy by
\be\label{can-S}
S_c^{HF} = \ln Z^{HF}_c +  \beta E_c^{HF} \;.
\ee
With these definitions, the average state density is given by Eq.~(\ref{sden}) with the canonical energy and entropy replaced by the their HF expressions (\ref{can-E}) and (\ref{can-S}). 
We note that the above $E_c^{HF}$ differs from the expectation value of $\hat H$ in the particle-number projected HF density $D^{HF}_N \equiv \hat P_N 
e^{-\beta (\hat H_{HF} - \lb \hat V \rb)}/ Z^{HF}_N$.  Similarly $S_c^{HF} \neq -\Tr (\hat D^{HF}_N \ln \hat D^{HF}_N)$.  The reason for these differences is the dependence of $\hat{H}_{HF}$ on the grand canonical one-body density $\varrho$.  

Similar relations define the approximate canonical energy and entropy in the finite-temperature HFB approximation starting from the particle-number projected proton (neutron) HFB partition function $Z^{HFB}_{N_{p(n)}}$ given by Eqs.~(\ref{HFB-partition}) and (\ref{z-HFB-tr}).  

As shown in Eqs.~(15) and (16) of Ref.~\cite{Alhassid2016}, one can reduce the grand-canonical partition function  to an approximate canonical partition function by a 2D saddle-point approximation and then compute the state density by the saddle-point approximation given in Eq.~(\ref{sden}).  However, if the particle-number fluctuation is small, the saddle-point approximation yields incorrect results.  This error originates in the fact that the particle number is a discrete variable rather than a continuous variable.  In Ref.~\cite{Alhassid2016}, the authors replaced the standard saddle-point approximation with the discrete Gaussian (DG) approximation, in which the quantity $\zeta$ (which appears as a pre-factor in the saddle-point  approximation) is given by their Eq.~(21).  Here we use the results of the two distinct versions of the DG approximation formulated in Ref.~\cite{Alhassid2016}.  We will denote by DG1 the approximation in which $\zeta$ given by Eq.~(21) of Ref.~\cite{Alhassid2016}, and by DG2 the approximation in which $\zeta$ is given by Eq.~(25) of Ref.~\cite{Alhassid2016} where the particle-number variances are calculated directly in the mean-field approximation.  Because of the implicit dependence of the grand canonical mean-field Hamiltonian on the proton and neutron chemical potentials, DG1 and DG2 can give different results.

The authors of Ref.~\cite{Alhassid2016} also identified a correction term in the canonical energy and entropy that arises from the saddle-point approximation with respect to $\beta$ (see Eqs.~(19) and (20) of Ref.~\cite{Alhassid2016}). This correction improves the accuracy of the discrete Gaussian approximations.  Here we compare the results from our exact particle-number projection method with those obtained in the discrete Gaussian approximations that include this correction term. We note the usual 3D saddle-point approximation is equivalent to the discrete Gaussian approximation without this correction term when the particle-number fluctuations are large (for both protons and neutrons). 

\section{Results}\label{results}

Here, we present results for the particle-number-projected finite-temperature mean-field theories in three heavy nuclei.  In Sec.~\ref{dy162} we discuss $^{162}$Dy, a typical strongly deformed  nucleus for which the appropriate mean-field theory is the HF approximation.  In Sec.~\ref{sm148} we present results for $^{148}$Sm, a typical spherical nucleus with a strong pairing condensate. Finally, in Sec~\ref{sm150} we discuss a transitional nucleus $^{150}$Sm, in which the pairing condensate is deformed.  In this last nucleus, we use the general projection formula derived in Sec.~\ref{tr-condensate}.  The PNP results are compared with the discrete Gaussian approximations and with the SMMC results~\cite{Alhassid2008,Ozen2013}. 

We employed a configuration-interaction (CI) shell model Hamiltonian that includes a one-body part and a two-body interaction. The valence single-particle model space  consists of the orbitals $0g_{7/2},1d_{5/2},1d_{3/2,}2s_{1/2},0h_{11/2},1f_{7/2}$ for protons and $0h_{11/2},0h_{9/2},1f_{7/2},1f_{5/2},2p_{3/2},2p_{1/2}, 0i_{13/2},1g_{9/2}$ for neutrons. The corresponding total number of single-particle states (including their magnetic degeneracy) is $N_s=40$ for protons and $N_s=66$ for neutrons. The one-body Hamiltonian and two-body interaction used for the above nuclei are discussed in Refs.~\cite{Alhassid2008,Ozen2013}.

\subsection{Particle-number projected HF for a strongly deformed nucleus: $^{162}$Dy}\label{dy162}

\begin{figure}[h!]
\includegraphics[width=0.5\textwidth]{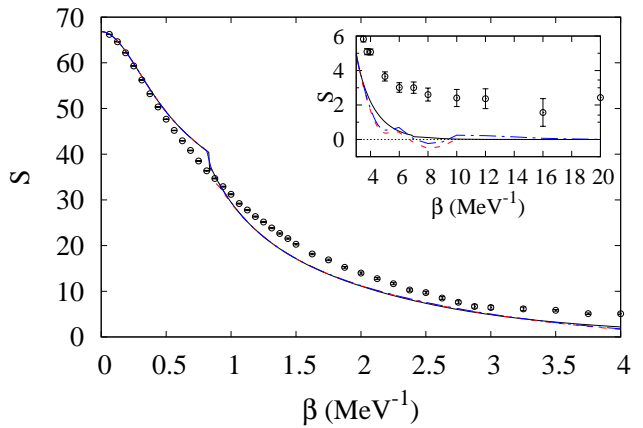}
\caption{\label{s_dy162}  Canonical entropy of $^{162}$Dy vs.~$\beta$ in the HF approximation. The approximate PNP canonical entropy (\ref{can-S})  (solid black line) is compared with the approximate canonical entropy from DG1 (dashed red line) and from DG2 (dashed-dotted blue line).  The open circles represent the SMMC entropy.  The inset shows the various entropies at higher $\beta$ values.}
\end{figure}

We applied the particle-number projected HF approximation to the strongly deformed nucleus $^{162}$Dy, in which the pairing is weak, using Eqs.~(\ref{HF-partition}) and (\ref{z-HF}) to calculate the particle-number projected partition function.  In Fig.~\ref{s_dy162} we compare the approximate canonical entropy (\ref{can-S}) from the PNP with those from the discrete Gaussian approximations and with the SMMC entropy.  The kink at $\beta \approx 0.83$ MeV$^{-1}$ in the HF results is due to the sharp shape transition that occurs in the grand canonical HF approximation.  At $\beta$ values below the shape transition, the HF results are in good agreement with those from the SMMC. However, at $\beta$ values above the shape transition, the entropies from the PNP and the discrete Gaussian approximations are noticeably lower than the SMMC entropy.  The reason for this discrepancy is that the HF approximation accounts only for the intrinsic $K$ states and not for the rotational bands that are built on top of these intrinsic states.  The PNP, DG1, and DG2 give nearly identical results for $\beta < 4$ MeV$^{-1}$.  In DG1 and DG2, however, the entropy exhibits unphysical oscillatory behavior for $ 4 \le \beta \le 10$ MeV$^{-1}$.  In contrast, the entropy asymptotes monotonically to zero at large $\beta$ in the PNP, as would be expected in a physical system. 

The PNP canonical excitation energy and state density closely resemble the corresponding results for the discrete Gaussian approximations, which are shown in Figs.~6 and 10 of Ref.~\cite{Alhassid2016}. The deviation between the PNP entropy and the discrete Gaussian entropy observed at low temperatures does not lead to any significant difference in the state densities.

The approximate canonical energy obtained from the derivative of the logarithm of the particle-number projected partition function [Eq.~(\ref{can-E})] is different from the canonical energy calculated from the expectation value of the Hamiltonian for the projected density $D_N^{HF}$.  However, we find these two values to be in very close agreement expect in the immediate vicinity of the shape transition.

\subsection{Particle-number projected HFB for a spherical condensate: $^{148}$Sm}\label{sm148}

\begin{figure}[h!]
\includegraphics[width=0.5\textwidth]{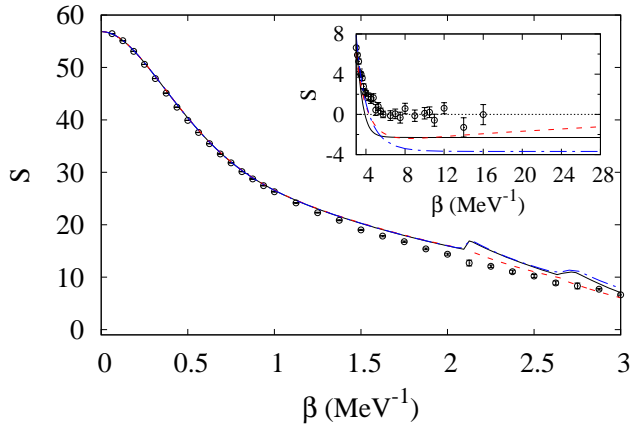}
\caption{\label{s_sm148}  Canonical entropies of $^{148}$Sm vs.~$\beta$ in the BCS limit of the HFB approximation.  Lines and symbols are as in Fig.~\ref{s_dy162}.  The inset shows the entropies for higher $\beta$ values.}
\end{figure}

To test our formulas for the particle-number projected HFB partition function in the BCS limit, given by Eqs.~(\ref{HFB-partition}) and (\ref{z-HFB-sph}), we apply the particle-number projected HFB approximation to $^{148}$Sm, for which the pairing condensate is spherical.  The canonical entropies for the PNP, the discrete Gaussian approximations, and the SMMC are shown in Fig.~\ref{s_sm148}.  The kinks in the HFB results in the region $2 \le \beta \le 3$ MeV$^{-1}$ are due to the proton and neutron pairing transitions.  For $\beta$ values below the first pairing transition, there is good agreement between the HFB results and the SMMC results.  The kinks that indicate the pairing transitions are more pronounced in the PNP and DG2 than in DG1.  

In the paired phase, the approximate canonical entropies decrease rapidly, dropping below zero around $\beta \approx 4$ MeV$^{-1}$.  A negative entropy is unphysical because the entropy of a non-degenerate ground state is zero.  This negative entropy originates in the intrinsic violation of particle-number conservation in the grand canonical HFB approximation and will be explained in detail for the PNP in Sec.~\ref{neg-entropy}.  An explanation for the discrete Gaussian approximations is given in Ref.~\cite{Alhassid2016}.  

However, for large $\beta$ values the PNP exhibits qualitative and quantitative improvements over the discrete Gaussian approximations.  The entropy from the PNP asymptotes smoothly to a value of $S_{c} \approx -2.30$.  The entropy from DG1 reaches a minimum around $\beta \approx 8$ MeV$^{-1}$ and subsequently increases with increasing $\beta$.  Such an increase is unphysical because, for these values of $\beta$, the system is already in its ground state. The entropy from DG2 asymptotes smoothly to a negative value of $S_{c} \approx -3.68$.  The absolute error in the estimate for the ground-state entropy is thus larger in DG2 than in the PNP by more than a unit in entropy.  

At large values of $\beta$, the PNP excitation energy for $^{148}$Sm closely resembles the DG1 excitation energy shown in Fig.~13 of Ref.~\cite{Alhassid2016}.  The PNP state density is similar to the DG2 state density shown by the dotted line in Fig. 16 of Ref.~\cite{Alhassid2016} while the DG1 state density is somewhat enhanced at low excitation energies.

\subsection{Particle-number projected HFB for a deformed condensate: $^{150}$Sm}\label{sm150}
To calculate the particle-number projected HFB partition function for $^{150}$Sm, which has a deformed pairing condensate, we must use the general PNP HFB formalism of Sec.~\ref{tr-condensate}, i.e., Eqs.~(\ref{HFB-partition}) and (\ref{z-HFB-tr}).  In this case, the advantages of the PNP over the discrete Gaussian approximations are significant for the excitation energy, canonical entropy, and state density.  In particular, DG1, the more accurate of the two discrete Gaussian methods used in Ref.~\cite{Alhassid2016}, becomes numerically unstable for all temperatures below the shape transition.  In contrast, the PNP remains stable for all temperatures.

\subsubsection{Excitation energy}

\begin{figure}[h!]
\includegraphics[width=0.5\textwidth]{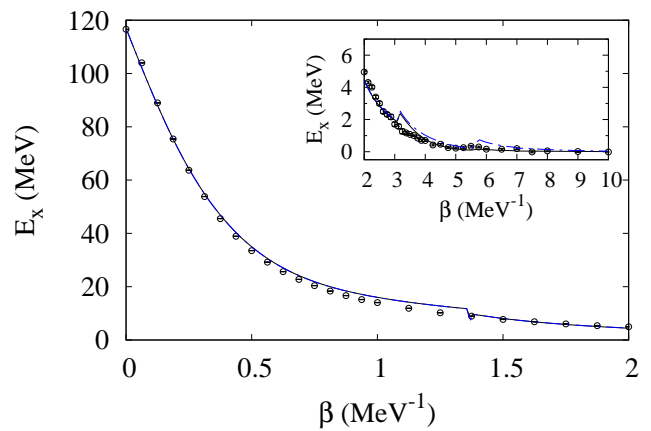}
\caption{\label{ex_sm150}  Excitation energy of $^{150}$Sm vs. $\beta$ in the HFB approximation.  The approximate canonical energy calculated from the PNP (solid black line) is compared with the approximate canonical energy from DG2 (dashed-dotted blue line).  DG1 becomes unstable for this nucleus and is not shown.   The open circles are the SMMC excitation energies. The inset shows higher $\beta$ values.}
\end{figure}

In Fig.~\ref{ex_sm150}, we show the excitation energy as a function of $\beta$ for the PNP and the discrete Gaussian approximation DG2 compared with the SMMC energy.  The kink in the HFB results $\beta \approx 1.5$ MeV$^{-1}$ is the shape transition, and the kinks at $\beta \approx 3$ MeV$^{-1}$ and $\beta \approx 6$ MeV$^{-1}$ are the proton and neutron pairing transitions.  Except around the phase transitions, there is good agreement between the PNP and DG2 results and the SMMC results.  DG2 shows a larger discrepancy from the SMMC around the pairing transitions than the PNP does.  DG1 (not shown in the figure) agrees with the SMMC and the other approximate canonical methods at low $\beta$ but breaks down at the shape transition and is not useful for $\beta$ values above the transition. 

\subsubsection{Canonical entropy}
\begin{figure}[h!]
\includegraphics[width=0.5\textwidth]{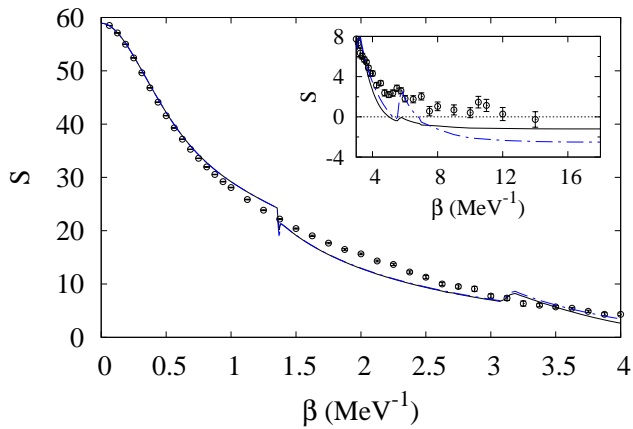}
\caption{\label{s_sm150}  Canonical entropy of $^{150}$Sm vs.~$\beta$ in the HFB approximation. Lines and symbols for the PNP, discrete Gaussian approximation DG2, and SMMC are as in Fig.~\ref{ex_sm150}.  The inset shows an expanded scale at large values of $\beta$.}
\end{figure}

The canonical entropy from the PNP, the discrete Gaussian approximation DG2, and the SMMC are shown in Fig.~\ref{s_sm150}.  At $\beta$ values below the shape transition, there is close agreement between the SMMC and the HFB results.  Between the shape transition at $\beta \approx 1.5$ MeV$^{-1}$ and the proton pairing transition at $\beta \approx 3$ MeV$^{-1}$, the PNP and DG2 entropies are lower than the SMMC entropy because of the contributions of rotational bands to the SMMC.  For $\beta$ values above the proton pairing transition, the HFB entropies are reduced even further, falling below zero at $\beta \approx 4$ MeV$^{-1}$.  As in the case of $^{148}$Sm, this negative entropy in the pairing phase originates in the violation of particle-number conservation in the HFB approximation.  

The entropy from the PNP is very close to that from DG2 in the unpaired phase but shows a quantitative improvement over DG2 in the paired phase. The entropy from the PNP asymptotes to $S_c \approx  -1.20$, while the entropy from DG2 asymptotes to $S_c \approx -2.52$.  As with $^{148}$Sm, the absolute error of the ground-state entropy in the PNP is lower by more than a unit of entropy than the error in DG2.  Furthermore, DG2 shows a large spike near the neutron pairing transition, which is not present in the PNP.

\subsubsection{State density}

\begin{figure}[h!]
\includegraphics[width=0.5\textwidth]{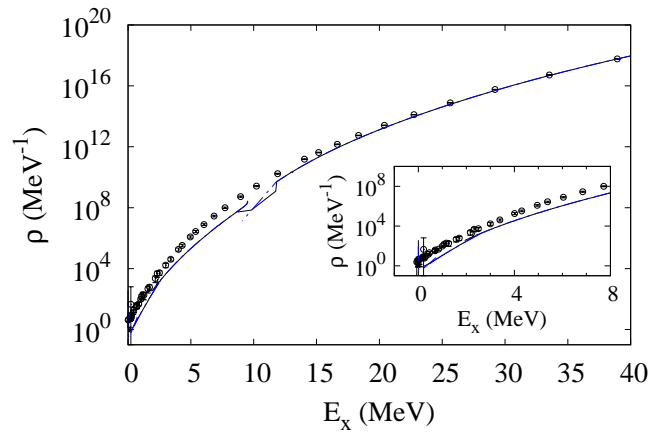}
\caption{\label{rho_sm150}  State density of $^{150}$Sm vs.~excitation energy $E_x$ in the HFB approximation.  Lines and symbols are as in Fig.~\ref{ex_sm150}.  The inset shows the low excitation energy results.}
\end{figure}

The behavior of the state density, which is shown for the PNP, the discrete Gaussian approximation DG2, and the SMMC in Fig.~\ref{rho_sm150}, closely resembles that of the canonical entropy.  At energies above the shape transition, the PNP and DG2 results agree well with the SMMC results. 
The HFB results are reduced at energies below the shape transition and reduced further at energies below the pairing transitions.  The discontinuities in both the PNP and in DG2 around $E_x \approx 10$ MeV and $E_x \approx 3$ MeV are due to the sharp kinks at the shape transition and proton pairing transition, respectively.  The discrepancy between the PNP and DG2 entropies in the high-$\beta$ limit is not noticeable in the state density because this discrepancy becomes significant only for very low excitation energies.

\subsection{Approximate canonical HFB entropy in the limit $T \to 0$}\label{neg-entropy}

As shown in Secs.~\ref{sm148} and \ref{sm150} for the nuclei with pairing condensates, $^{148}$Sm and $^{150}$Sm, the approximate canonical entropy from the PNP asymptotes to a negative number.  Here, we show how this unphysical result is due to the inherent violation of particle-number conservation in the grand canonical HFB approximation.  We consider the limit of sufficiently large $\beta$, in which we can neglect the contribution of excited states to the partition function. In this limit,
\be \label{zerotlim}
Z^{HFB}_{N_{p(n)}} \to e^{-\beta E_0}\bra{\Phi}\hat{P}_{N_{p(n)}}\ket{\Phi} \;,
\ee
where $E_0$ is the ground-state energy and $\ket{\Phi}$ is the HFB ground state.
This state can be written as a linear superposition of states with even particle numbers,
\be
\ket{\Phi} = \sum_{N=0,2,4,...}\alpha_N\ket{\psi}_N \;,
\ee
where $\ket{\psi}_N$ is an $N$-particle state.  Eq.~(\ref{zerotlim}) can then be expressed as
\be
Z^{HFB}_{N_{p(n)}} \to e^{-\beta E_{0}}\abs{\alpha_{N_{p(n)}}}^{2} \;.
\ee
where $\abs{\alpha_{N_{p(n)}}}^{2} $ is the probability that the HFB ground-state condensate contains $N_{p(n)}$ particles.  Because particle-number is not conserved, $\abs{\alpha_{N_{p(n)}}}^{2} < 1$.  Using Eq.~(\ref{can-S}), we find in the limit of zero temperature 
\be
S^{HFB}_c \to \ln\abs{\alpha_{N_p}}^{2} + \ln\abs{\alpha_{N_n}}^{2}\;.
\ee
Since $\abs{\alpha_{N_{p(n)}}}^{2} < 1$, this entropy is the sum of two negative numbers and is therefore negative. A closely related explanation for the discrete Gaussian approximations is given in Ref.~\cite{Alhassid2016}.  This negative entropy is an inherent limitation of the particle-number projection after variation method in the grand canonical HFB theory.  

\subsection{Computational Efficiency}

Another advantage of the PNP over the DG1 approximation is its computational efficiency.  We emphasize DG1 because this approximation was shown in Ref.~\cite{Alhassid2016} to be more accurate than DG2 for $^{162}$Dy and $^{148}$Sm.

Both the PNP and the discrete Gaussian methods require finding the self-consistent mean-field solutions for a set of $\beta$ values.  The additional cost of the PNP HF approximation scales as $N_s^2$, since calculating $\zeta_n^{HF}$ in Eq.~(\ref{z-HF}) takes $N_s$ operations and must be done for each of the $N_s$ quadrature points in the Fourier sum.  The PNP HFB approximation scales as $N_s^4$, because the matrix decomposition in the stabilization method (discussed in Appendix II) requires $N_s^3$ operations for each of the $N_s$ quadrature points.  

In the discrete Gaussian approximation DG1, it is necessary to calculate numerically the derivatives of the proton and neutron numbers with respect to the chemical potentials, which requires finding additional mean-field solutions.  The cost of calculating these derivatives accurately can be large, especially in the vicinity of the phase transitions where the mean-field solution can take many iterations to converge.  

\section{Conclusion and outlook}\label{conclusion}

We have derived expressions for exact particle-number projection after variation in the finite-temperature mean-field approximations. 
 In the HF approximation, the PNP is straightforward because particle number is conserved.  However, in the HFB approximation, the violation of particle-number conservation in the paired phase makes the PNP more difficult.  We have derived an unambiguous expression for the particle-number projected HFB partition function under the assumption that the condensate is time-reversal invariant. In the BCS limit, and, in particular, for a spherical condensate, this expression simplifies considerably.  

We have assessed the accuracy of the PNP in both the HF and HFB approximations, using the SMMC as a benchmark.  In addition, we have compared the performance of this method to the discrete Gaussian approximations formulated in Ref.~\cite{Alhassid2016}, which usually improve the saddle-point approximation.  Our results show that, as is the case with the discrete Gaussian approximations, the PNP is in agreement with the SMMC for temperatures above the shape or pairing phase transitions.  In general, we find that the PNP provides both quantitative and qualitative improvements over the discrete Gaussian approximations at low temperatures.  In the HF case, the PNP suppresses an instability that develops in the canonical entropy calculated by the discrete Gaussian approximations at large $\beta$.  In the paired phase of the HFB, the PNP entropy shows the correct qualitative behavior, i.e., it is monotonically decreasing with increasing $\beta$, unlike the DG1 approximation.  In this paired phase, the PNP reduces the error in the DG2 approximation by more than a unit of entropy.  This reduction is significant, since the errors in both cases are of order unity.  Finally, the PNP is significantly more computationally efficient than DG1, which was shown in Ref.~\cite{Alhassid2016} to be the more accurate of the two discrete Gaussian approximations.

However, the projection after variation method is inherently limited by the grand canonical mean-field theories to which it is applied.  In a deformed nucleus, the PNP mean-field theory cannot describe the rotational enhancement that is observed in the SMMC below the shape transition temperature. In a nucleus with a strong pairing condensate, the intrinsic violation of particle-number conservation below the pairing transition temperature will necessarily lead to a negative ground-state value of the PNP canonical entropy.  It is therefore desirable to explore improvements to finite-temperature mean-field theories that can address these issues.

One avenue for improvement is the variation after projection (VAP) method.  VAP would involve a considerable computational cost.  In particular, the entropy of the projected HFB density $S_N = -\Tr(\hat{D}_N \ln \hat{D}_N)$, which is required for the calculation of the free energy at finite temperature, is difficult to calculate in the paired phase since the HFB Hamiltonian does not commute with the particle-number operator.  VAP conserves particle-number during the variation and is therefore expected to smooth the sharp transitions of the mean-field approximations and the unphysical negative entropy in the paired phase.  Given the potential advantages of this method, it would be worth investigating the possibility of developing a VAP method for the finite-temperature HF and HFB approximations.

Another improvement would be to use the PNP in the static-path approximation (SPA)~\cite{Muhlschlegel1972, Alhassid1984, Lauritzen1988,Puddu1991}.  The SPA takes into account the static fluctuations of the mean field  beyond its self-consistent solution.  As with the mean-field approximation, the particle-number projection can be carried out either before or after the SPA integration.  It would be useful to conduct a systematic assessment of such a particle-number projected SPA.

Finally, the PNP expressions derived here for the HFB approximation assume that the condensate is time-reversal invariant and are thus not completely general.  For example, the condensates of odd-even and odd-odd nuclei break time-reversal symmetry.  A general PNP formula was derived in Ref.~\cite{Rossignoli1994}, but the application of this formula is limited by a sign ambiguity.  For the case of statistical density operators, this sign ambiguity can be resolved by relating the trace over the Fock space to a pfaffian~\cite{Robledo2009}.  In a forthcoming publication, we will show how to extend this pfaffian approach to the traces involved in the particle-number projection at finite temperature~\cite{Fanto-tobepublished}.

\section*{Appendix I: Trace of exponential of fermionic one-body operator}

Here we derive the formula for the grand canonical trace of the exponential of a quadratic fermionic operator. Assuming a diagonalizable $N_{s} \times N_{s}$ matrix $\mathcal{C}$, 
\be\label{expc}
 \Tr \,e^{\xi^\dagger {\cal C} \xi} = \det\left(1 + e^{\mathcal{C}}\right) \;,
\ee
where $\xi$ is defined in Sec.~\ref{tr-condensate}. 

Since $\mathcal{C}$ is diagonalizable, there is a similarity transformation $S$ that bring $\mathcal{C}$ to a diagonal form
\be
S\mathcal{C}S^{-1} = \mathcal{D} = \left(\begin{matrix}
 \lambda_{k_{1}} & & & & & \\
 & \ddots & & & & \\
 & & \lambda_{k_{N_{s}/2}} & & &\\
 & & & \lambda_{\bar{k}_{1}} & &  \\ 
 & & & & \ddots & \\
 & & & & & \lambda_{\bar{k}_{N_{s}/2}} \\
\end{matrix}\right) \;.
\ee
If $S$ is unitary, it can be used to transform $\alpha_k$ and $\alpha^\dagger_k$ to new fermionic operators $d_k$ and $d^\dagger_k$ such that
\be\label{trace-d}
\begin{split}
 \Tr \,e^{\xi^\dagger {\cal C} \xi}& = \Tr\, e^{ \sum_{k >0} (\lambda_k d^\dagger_k d_k + \lambda_{\bar k}  d_{\bar k} d^\dagger_{\bar k})} \\
 &=  \Tr\, e^{ \sum_{k >0} (\lambda_k d^\dagger_k d_k - \lambda_{\bar k} d^\dagger_{\bar k} d_{\bar k} + \lambda_{\bar k} )} \;.
 \end{split}
\ee
Since $d_k$ are fermionic operators, the grand canonical trace in Eq.~(\ref{trace-d}) can be evaluated as usual to give
\be\label{trace-d1}
\begin{split}
 \Tr \,e^{\xi^\dagger {\cal C} \xi} & = \prod_{k>0} \left[(1+e^{\lambda_k})(1+e^{-\lambda_{\bar k}}) e^{\lambda_{\bar k}} \right] \\
 & = \prod_{k>0} \left[(1+e^{\lambda_k})(1+e^{\lambda_{\bar k}})\right] = \det \left(1+e^{\mathcal{C}} \right)\;.
 \end{split}
 \ee
However, in general, $S$ is not unitary.  In this case, we make use of a nonunitary Bogoliubov transformation \cite{Balian1969}.  We define the operators $\{d,\tilde{d}\}$ by the canonical transformation
\be\label{can}
\eta = S\xi\;, \qquad \tilde{\eta} = \xi^\dagger S^{-1} \;,
\ee
where $\eta = \left(d_{k_{1}},\ldots,d_{k_{N_{s}/2}}, \tilde d_{\bar{k}_{1}},\ldots,\tilde d_{\bar{k}_{N_{s}/2}}\right)^T$ and $\tilde \eta =  \left(\tilde d_{k_{1}},\ldots,\tilde d_{k_{N_{s}/2}}, d_{\bar{k}_{1}},\ldots, d_{\bar{k}_{N_{s}/2}}\right)$.  It can be shown that $\{d,\tilde{d}\}$ have the same anti-commutation relations as $\{\alpha,\alpha^{\dagger}\}$. 
However, this transformation does not preserve the hermiticity relation of the operators.
Therefore we must treat the kets and bras related to these operators differently.  Because the anti-commutation relations are preserved, the usual creation and annihilation formulas apply to the left and right bases separately.  We define the left and right vacuums by
\be
\ket{0}_{d} = \prod_{k > 0} d_{k}d_{\bar{k}}\ket{0}, \, _{d}\bra{\bar{0}} = \bra{0}\prod_{k > 0}\tilde{d}_{k}\tilde{d}_{\bar k} \;,
\ee
and left and right states by
\be
\begin{split}
\ket{\phi} &= \prod_{k>0}(\tilde{d}_{k})^{n_{k}}(\tilde{d}_{\bar k})^{n_{\bar k}}\ket{0}_{d}\\
\bra{\bar{\phi}} & = _{d}\bra{\bar{0}} \prod_{k>0} (d_{k})^{n_{k}}(d_{\bar k})^{n_{\bar k}} \;.
\end{split} 
\ee
The anti-commutation relations ensure that
\be
\braket{\bar{\phi}}{\phi^\prime} = \delta_{\phi,\phi^\prime}, \,\,\bra{\bar{\phi}}\tilde{d}_{k}d_{k}\ket{\phi} = n_{k} \;,
\ee
where $n_{k}$ is the occupation number of state $k$.  Furthermore, as discussed in Ref.~\cite{Balian1969}, these left and right states form a a bi-orthogonal basis for the Fock space and therefore satisfy the completeness relation, 
\be
\sum_{\phi}\ket{\phi}\bra{\bar{\phi}} = 1 \;.
\ee
We can rewrite $\xi^\dagger\mathcal{C}\xi$ in this bi-orthogonal basis
\be
\xi^\dagger\mathcal{C}\xi = \tilde{\eta}S\mathcal{C}S^{-1}\eta = \sum_{k > 0} \lambda_k \tilde{d}_{k}d_{k} - \lambda_{\bar k}\tilde{d}_{\bar k}d_{\bar k} + \lambda_{\bar k} \;.
\ee

We can now compute the trace as follows ($\ket{\psi}$ below is an arbitrary state in the $\alpha,\alpha^\dagger$ basis):
\be
\begin{split}
 \Tr \,e^{\xi^\dagger {\cal C} \xi}  & = \sum_{\psi}\bra{\psi}e^{\xi^\dagger {\cal C} \xi}\ket{\psi} \\ & = \sum_{\psi,\phi,\phi^{\prime}} \braket{\psi}{\phi}  \bra{\bar{\phi}}e^{\sum_{k > 0} \left(\lambda_k \tilde{d}_{k}d_{k} - \lambda_{\bar k}\tilde{d}_{\bar k}d_{\bar k} + \lambda_{\bar k}\right)}\ket{\phi^{\prime}} \braket{\bar{\phi^\prime}}{\psi} \\
& = \sum_{\phi}\bra{\bar{\phi}}\prod_{k > 0}e^{\lambda_k \tilde{d}_{k}d_{k} - \lambda_{\bar k}\tilde{d}_{\bar k}d_{\bar k} + \lambda_{\bar k}}\ket{\phi} \\
& = \prod_{k > 0}\sum_{\substack{n_{k} = 0,1 \\ n_{\bar{k}} = 0,1}}e^{\lambda_k n_{k} + \lambda_{\bar {k}}(1-n_{\bar{k}})} \\
& = \prod_{k > 0} \left(1+ e^{\lambda_{k}}\right)\left(1+e^{\lambda_{\bar{k}}}\right)  = \det\left(1 + e^{\mathcal{C}}\right) \;.
\end{split}
\ee
This completes the proof of Eq.~(\ref{expc}). 

\section*{Appendix II: Stabilization of the HFB particle-number projection}

Eq.~(\ref{z-HFB-tr}) becomes numerically unstable at large values of $\beta$. We rewrite
\be
\begin{split}
\det\left(1 +  \mathcal{W}^\dagger e^{i\phin\mathcal{N}}\mathcal{W}e^{-\beta\mathcal{E}} \right)  = &\det\left(\mathcal{W}^\dagger\right)\det\left( e^{i\phin\mathcal{N}}\right)\det\left(\mathcal{W}\right)\\&\times\det\left(\mathcal{W}^\dagger e^{-i\phin\mathcal{N}}\mathcal{W} + e^{-\beta\mathcal{E}}\right).
\end{split}
\ee
Using $\det\left(e^{i\phin\mathcal{N}}\right) = 1$ and $\det\mathcal{W}^\dagger = \left[\det\mathcal{W}\right]^{-1}$, we find
\be
\det\left(1 +  \mathcal{W}^\dagger e^{i\phin\mathcal{N}}\mathcal{W}e^{-\beta\mathcal{E}} \right) = \det\left(\mathcal{W}^\dagger e^{-i\phin\mathcal{N}}\mathcal{W} + e^{-\beta \mathcal{E}}\right).
\ee
 The determinant on the r.h.s.~can be computed stably~\cite{Loh1992} by decomposing the matrix $A_n \equiv \mathcal{W}^\dagger e^{-i\phin\mathcal{N}}\mathcal{W} + e^{-\beta \mathcal{E}}$ in the form
\be
A_n = Q_n D_n R_n \;,
\ee
where $Q_n$ is an orthogonal matrix, $R_n$ is an upper triangular matrix in which each diagonal entry is $1$, and $D_n$ is a diagonal matrix.  $Q_n$ and $R_n$ are well-conditioned matrices, while $D_n$ contains the scales of the problem.  Consequently, the eigenvalues of $Q_n$ and $R_n$ can be computed stably.  We use these eigenvalues, together with the diagonal entries of $D_n$, to find $\det A_n$.  In practice, to avoid numerical overflow, we compute the quantity
\be
\ln \det A_n = \ln \det Q_n + \ln \det D_n + \ln \det R_n \;.
\ee

\end{document}